\documentclass[aps,prb,twocolumn,groupedaddress]{revtex4}
\usepackage{graphicx}% Include figure files
\usepackage{dcolumn}% Align table columns on decimal point

\begin{document}

%\preprint{}

\title{Structure and stability of Ba-Cu-Ge type-I clathrates}


\author{Yang Li}
\author{Ji Chi}
\author{Weiping Gou}
\author{Sameer Khandekar}
\author{Joseph H. Ross, Jr.}
%\thanks{}
\affiliation{Department of Physics, Texas A\&M University, College Station, TX 77843-4242}


\date{\today}

\begin{abstract}
We have prepared samples of nominal type Ba$_{8}$Cu$_{x}$Ge$_{46-x}$ by 
induction melting and 
solid state reaction.  Analysis shows that these materials form type-I 
clathrates, with copper content between $x$ = 4.9 and 5.3, nearly independent 
of the starting composition.  We used x-ray powder diffraction and single-crystal 
electron diffraction to confirm the cubic type-I clathrate structure, 
while electron microprobe measurements 
confirmed the stability of the $x \approx$ 5 composition.  This result 
differs from the corresponding Ag and Au clathrates, and was not 
known previously due perhaps to the similar Cu and Ge form factors in
x-ray diffraction.  The observed composition adheres very tightly
to a valence-counting scheme, in agreement with a Zintl-type stability
mechanism.  This implies a gap in the electronic density of states,
also in contrast to the metallic behavior of the Au and Ag analogs.
Magnetization measurements showed a large diamagnetic response
in the Ba-Cu-Ge clathrate.  
This behavior is consistent with semiconducting or semimetallic
behavior, and is similar to that of a number of
intermetallic semiconductors.
\end{abstract}

%\keywords{}

\maketitle

%\section{}
%\subsection{}
%\subsubsection{}
\section{Introduction}

Silicon, germanium, and tin form clathrate phases having framework 
structures in which the cages enclose single ions of the alkali 
metal, alkali earth, rare earth or halogen 
series.\cite{ramach99,nolasPR00,nolasCM00}
Doping with metal
atoms has led to a wide variety of electronic behavior, including 
superconductivity in Ba$_{8}$Si$_{46}$,\cite{yamanaka00} and in
Ba$_{8}$Ga$_{16}$Ge$_{30},$\cite{bryan99} 
ferromagnetism in Ba$_{8}$Mn$_{2}$Ge$_{44}$,\cite{kawaguchi00} and a 
number of semiconducting compositions including the silicon-only 
material
Si$_{136}$.\cite{gryko00} There has been considerable interest in
these materials, for a number of potential applications including
thermoelectric cooling.\cite{nolasPR00,nolasCM00,cohn99}
There is the further potential for epitaxial growth in conventional 
semiconductor systems,\cite{munetoh01} and transition-metal and 
rare-earth-doped
clathrates offer the possibility of new magnetic semiconductors.  

We have studied the formation of Ba-Cu-Ge clathrates, and report on
the stable formation of clathrates with the type-I structure and
compositions close to Ba$_{8}$Cu$_{5}$Ge$_{40}$.  This composition
forms preferentially for a range of starting compositions, for 
ambient-pressure synthesis.  The formation of Ba-Cu-Ge type-I
clathrates has been reported previously.\cite{cordier91}  It is known
that Cu substitutes for Ge on framework sites, although the
nearness of Cu and Ge in the periodic table makes identification
of specific Cu atomic positions via x-ray analysis
somewhat difficult.

In type-I germanium-barium clathrates, of nominal composition 
Ba$_{8}$Ge$_{46}$, germanium
occupies three sites, which are Wyckhoff 6$c$, 16$i$, and 24$k$ 
sites in the cubic Pm$\bar 3$n (\#223) structure.  This structure
is the analog of the 
(Cl$_{2}$)$_{8}$(H$_{2}$O)$_{46}$ gas clathrate.\cite{nolasPR00}
Ba ions occupy 2$a$ and 6$d$ sites, corresponding to locations
within the two distinct
cages of the Ge framework.  A stable Ba-Ge type-I clathrate of
composition Ba$_{8}$Ge$_{43}$, with three framework vacancies
per unit cell, has been prepared and found to
be semiconducting.\cite{herrmann99} The analogous Ba-Sn 
clathrate is predicted to be stable in the composition 
Ba$_{8}$Sn$_{42}$.\cite{myles01} The latter can be understood
in terms of the Zintl-Klemm concept,\cite{Kauzlarich96} 
assuming that each Ba donates two electrons to the framework,
satisfying the tetrahedral bonding requirements of the framework
with four electrons per site, including vacancies.
Semiconducting Ba$_{8}$Ge$_{43}$ has one less vacancy per
cell than given by this argument, perhaps due to a slightly
smaller electron transfer from Ba.\cite{herrmann99}

A number of transition metals have been incorporated into
Ge and Si clathrates,\cite{cordier91} predominantly by substitution 
at the 6$c$ site. Au and Ag-doped Ba-Si clathrates have recently been
reported with a range of doping levels, exhibiting metallic
and superconducting behavior.\cite{herrmann98,nozue00} Thus, 
these materials must lie outside the range of stabilization via bond
filling, of the Zintl type as seen in other clathrates.  
Au-doped Ba-Ge clathrate has also been reported to be 
metallic.\cite{herrmann99} Hence by analogy we expected the
Cu-doped Ba-Ge clathrate to be a metal with a range of 
stable composition,
although the initial report described only 
Ba$_{8}$Cu$_{6}$Ge$_{40}$.\cite{cordier91}
However, our results show that Cu prefers to
substitute in a relatively narrow range of composition, as 
detailed below.  Furthermore, magnetic and transport measurements
indicate this material to be semimetallic, in contrast to the
Au analog, and in the Analysis and Discussion section we 
show that a bond filling mechanism provides a good accounting
for the stability of this phase.  

\section{Experiment}

Ingots with nominal compositions 
Ba$_{8}$Cu$_{x}$Ge$_{46-x}$, for 
$x$ = 2, 4 and 6, were formed by rf induction melting and 
solid-state reaction. These samples will be referred 
to as Cu2, Cu4, and Cu6, respectively.
Stoichiometric quantities of the elemental materials were finely
powdered and pressed into BN crucibles, then induction 
heated in an argon atmosphere. The resulting ingots, exhibiting a metallic luster,
were further reacted at $950^{\circ}$ C for 3 days followed by 
$700^{\circ}$ C for 4 days, in evacuated ampoules. SEM and EDS 
analysis for the Cu2 
and Cu6 samples showed these ingots to consist of 
large clathrate crystallites of typical size several-100 $\mu$m, with 
smaller crystallites of other phases. The outer portion of these 
ingots was physically separated, and the central portion reserved
for x-ray and magnetization measurements, giving a sample somewhat
more concentrated in clathrate.  Microprobe measurements were carried
out on samples of the entire ingots.

\begin{figure*}
\includegraphics{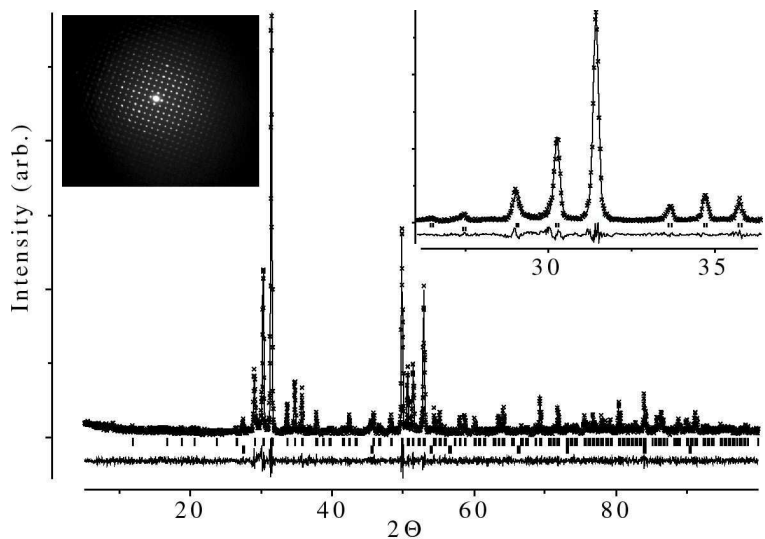}%
\caption{\label{fig:fig1}Cu K-$\alpha$ powder x-ray pattern 
for the Cu2
sample (nominal composition Ba$_{8}$Cu$_{2}$Ge$_{44}$),
with fit and difference plot.
Inset at upper right is an expanded view.  
Vertical bars below the spectra indicate positions of
clathrate-structure and Ge reflections.
The inset at left shows an electron diffraction pattern 
for the Cu6 sample,
showing a cubic pattern characteristic of the type-I clathrate,
with no superlattice structure.}
\end{figure*}

Analysis by Cu $K\alpha$ powder x-ray diffraction showed
characteristic type-I clathrate reflections,
with a few percent diamond-structure Ge in the Cu2 and
Cu4 samples. LeBail extraction and Rietveld refinement were 
performed using
GSAS software,\cite{gsas,expgui} giving lattice constants
steadily decreasing with increasing nominal Cu concentration.
The refinement process is rather insensitive
to the difference between Cu and Ge,
due to their nearness 
in the periodic table. Therefore, for better 
determination
we used electron microprobe results, described
below, to constrain the Cu/Ge ratio in the clathrate
during the x-ray
fitting process. Furthermore, in the fit we assumed
the Cu to be located only on the 6$c$ site, as reported
for other transition-metal Ge clathrates.
Fig.~\ref{fig:fig1} shows an x-ray
fit thus obtained. 

\begin{figure}
\includegraphics{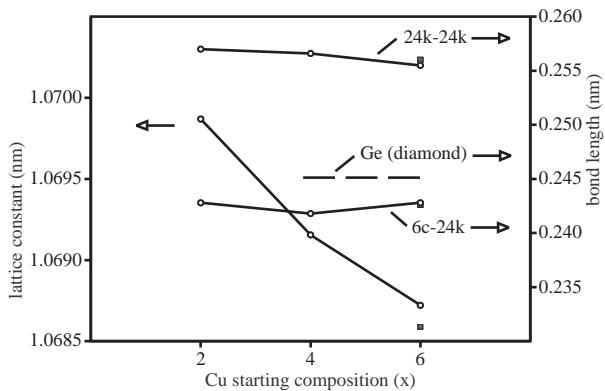}%
\caption{\label{fig:fig2}Lattice constant and 
representative framework bond lengths
versus nominal Cu composition.  
Near-neighbor bond length for diamond-structure
Ge is shown by the dashed line.
The shaded squares are from 
Ref.~\onlinecite{cordier91}, for a sample of 
starting composition $x$=6.}
\end{figure}

X-ray analysis results are summarized in 
Table~\ref{tab:table1},
with the trend in some of the structural parameters
also plotted in Fig.~\ref{fig:fig2}.
We found that the Cu starting composition 
had a rather small effect on the resulting Cu and vacancy
concentrations in the clathrate, with the Cu content
very close to 5 per cell in all cases.  The lattice constant
exhibited a small decrease with increasing Cu
content (Fig.~\ref{fig:fig2}).  These results are consistent with those 
of Ref.~\onlinecite{cordier91}; values from
that work are shown as filled squares in 
Fig.~\ref{fig:fig2}, placed at $x$=6, which was 
the starting composition for that study.  Bond
lengths in Fig.~\ref{fig:fig2} show some sample-to-sample
variation. Error bars are smaller than this
variation.  We used isotropic thermal parameters for
these fits, and found the largest thermal parameter
(U$_{iso} \approx 0.03$) for the Ba 6$d$ site, 
occupying the larger cage, consistent with previous work on
Ge clathrates.

\begin{table}
\caption{\label{tab:table1}Measured parameters for the three 
samples obtained
from powder x-ray diffraction.  6$c$ angle is the larger 
24$k$-6$c$-24$k$ bond angle.}
\begin{ruledtabular}
\begin{tabular}{lccc}
Sample&\mbox{Cu2}&\mbox{Cu4}&\mbox{Cu6} \\ 
\hline
$a$ (nm)&1.06987(1)&1.06916(2)&1.06872(1) \\
$x$&0.1829&0.1835&0.1830 \\
$y$&0.3139&0.3150&0.3140 \\
$z$&0.1201&0.1200&0.1196 \\
6$c$ angle&110.2$^{\circ}$&109.8$^{\circ}$&109.9$^{\circ}$\\
\hline
occupancy:& & & \\
Ba 2$a$&1.00&1.00&0.93 \\ 
Ba 6$d$&1.00&0.99&0.97 \\
Ge 6$c$&0.09&0.10&0.10 \\ 
Cu 6$c$&0.82&0.81&0.88 \\
Ge 16$i$&1.00&0.99&0.97 \\ 
Ge 24$k$&1.00&1.00&1.00 \\ 
\hline
Ba/cell&8.0&8.0&7.7 \\ 
Cu/cell&4.9&4.9&5.3 \\ 
Ge/cell&40.6&40.5&40.1 \\ 
Cu+Ge vacancy&0.49&0.59&0.61 \\ 
\end{tabular}
\end{ruledtabular}
\end{table}

The parameters $x$, $y$, and $z$ in Table~\ref{tab:table1} 
are those for the Pm$\bar 3$n (\#223) structure;
the 16$i$ position is given by $(x, x, x)$, and the 
24$k$ position, $(0, y, z)$.  These parameters
change very little with composition, and the 6$c$ site 
remains close to perfect tetrahedral symmetry 
(bond angles 109.5$^\circ$). Fig.~\ref{fig:fig2} also
shows the range of framework bond lengths, as the
6$c$-24$k$ and 24$k$-24$k$ bonds are the shortest
and longest Ge-Ge bonds, respectively.

A powdered specimen
from the Cu6 ingot placed on a grid for
transmission-electron 
microscopy showed the characteristic cubic
diffraction pattern (Inset of fig.~\ref{fig:fig1}). 
Diffraction
images from several crystallites in this sample were
compared to simulations, giving a good match
to the observations. A superlattice structure
was previously identified in 
Ba$_{8}$Ge$_{43}$, attributed to preferential
ordering of the framework vacancies in 
that material.\cite{herrmann99}
We saw no evidence of a superlattice pattern in our
samples, indicating the Cu atoms and vacancies to
be randomly distributed on the 6$c$ and 16$i$ sites.

Compositional analysis was carried out
using a commercial wavelength dispersive spectroscopy 
(WDS)-based electron microprobe apparatus and 
analysis system (Cameca). In these measurements 
the clathrate phase was identified in each 
sample, plus Ge,  an oxide of approximate
composition BaGeO$_{2}$, and
Ge$_{3}$Cu$_{5}$ (Cu6 sample only).  Although dispersed
throughout the ingots, the latter two
were below the x-ray detection limit.
In the Cu4 sample a single localized
Ge-Ba-S phase was detected, attributed
to an impurity left after H$_{2}$SO$_{4}$ 
cleaning of the
mortar and pestle, however no S was detected in any of
the other analyses. The Ba-Ge oxide was not found
to contain Cu in any of the samples. 

X-ray and WDS analysis both provide a relative
measure of the 
number of atoms per cell, rather than an absolute
composition.  We produced the compositions in
Table~\ref{tab:table1} by assuming the largest 
site occupancy to be 100\%.  Scaling these results
to lower values results quite quickly in vacancy
concentrations which are unreasonable.  For the
Cu2 and Cu4 samples, this yields essentially
100\% occupancy of the Ba sites, with framework
vacancies appearing only on the 6$c$ site. 
The Cu6-sample results are somewhat different,
with vacancies appearing on the Ba 2$a$ and
framework 16$i$ sites, as well as the framework 
6$c$ site.
It appears that overloading the framework with Cu
atoms forces vacancies to appear on the 16$i$
site.  The presence of Ba 2$a$ vacancies is 
consistent with the behavior observed in
Ba$_{8}$Si$_{46}$.\cite{kitano01}

In the Cu2 and Cu4 samples, diamond-structure Ge 
appeared as inclusions
within the abundant clathrate phase, indicating a
phase separation during the solid state reaction
process.  
For Cu6, an additional phase 
was observed, of nominal composition GeCu$_{1.65}$
or Ge$_{3}$Cu$_{5}$. (Diamond-Ge also contained about 
1 at.\% Cu.)  Since the Cu content for Cu6 exceeded 
the clathrate stability limit, approximately 10\% of the Cu 
appeared as
Ge$_{3}$Cu$_{5}$ (or about 2 at.\% of the entire
sample), over and above the small change in Cu
content of the clathrate. 
Ge$_{3}$Cu$_{5}$ has not been 
previously reported
to our knowledge,\cite{pearson} and
this structure may be stabilized by
the adjacent phases.  None of the reported Ge-Cu 
intermetallic structures (GeCu$_{3}$, GeCu$_{5}$,
Ge$_{2}$Cu$_{5}$) could be identified in the x-ray
spectrum for the sample, nor could we identify 
reflections due to any other structures, within
our resolution.

\begin{figure}
\includegraphics{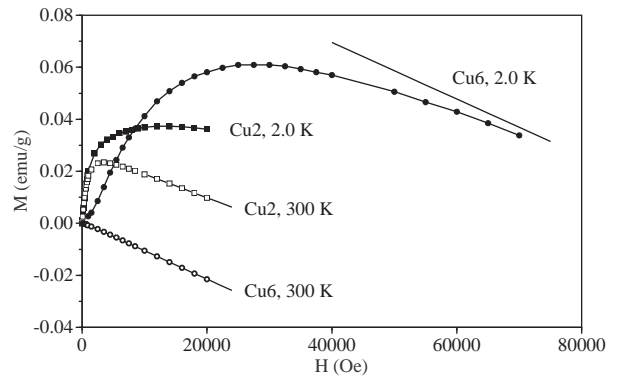}%
\caption{\label{fig:fig3}$M$ vs. $H$ results for the 
Cu2 and Cu6 samples, at 2.0 K and 300 K as labeled.  
Error bars are smaller than the symbols.
Solid lines through the 300 K data are least-squares
fits, giving $\chi _{dia} = -10.94(4) \times 10^{-7}$ 
emu/g (Cu6) and
$-9.02(4) \times 10^{-7}$ emu/g (Cu2).  
Solid curve above the Cu6, 2 K data
has the same slope as the 300 K fit, showing slow
saturation to this line.  Curves connecting 2.0 K
data are guides to the eye.}
\end{figure}

Fig.~\ref{fig:fig3} shows $M$ vs. $H$ results
for the Cu2 and Cu6 samples, obtained using a
commercial SQUID magnetometer
(Quantum Design, Inc.).  The Cu6 sample 
exhibited diamagnetism at high temperatures,
which we fit to $\chi _{dia} = -10.94(4) \times 10^{-7}$ 
emu/g (solid line in Fig.~\ref{fig:fig3} for 300 K).
At low temperatures the Cu6 sample also shows low-moment
ferromagnetism, which we attribute to a dilute concentration 
of random defects.\cite{liunpub}
However, the high-field
trend for the 2 K data is toward the same diamagnetic slope as at 
300 K (upper line in Fig.~\ref{fig:fig3}), though
some moments are not 
completely saturated in the 70 kOe maximum
field of our magnetometer.  The Cu2 sample
has a weak ferromagnetic phase visible at 300 K, but above
saturation the slope was fit to 
$\chi _{dia} = -9.02(4) \times 10^{-7}$ 
emu/g (solid line), very similar to that of the other
sample.  Thus we identify
the Cu-Ge clathrate to exhibit bulk diamagnetism with
a susceptibility of approximately $-10 \times 10^{-7}$ 
emu/g.

\section{Analysis and Discussion}

The hypothetical empty-cage clathrates Si$_{46}$, 
Ge$_{46}$, and Sn$_{46}$ are 
semiconducting as predicted by
electronic-structure
calculations.\cite{zhao99,myles01,saito95}
In filled Sn clathrates, the stable phase
Cs$_{8}$Sn$_{44}$ has two vacancies per cell, 
in accordance with the classic Zintl 
concept,\cite{Kauzlarich96}
in which the number of valence electrons on the
framework is preserved to give the most favorable
bonding configuration, and hence a filled band
and semiconducting behavior.  Ternaries such 
as Cs$_{8}$Ga$_{8}$Sn$_{38}$ 
and Cs$_{8}$Zn$_{4}$Sn$_{42}$
can also clearly be treated this way.\cite{myles01}

The stable Ba-Ge phase 
Ba$_{8}$Ge$_{43}$\cite{herrmann99,carrillo00}
can be treated as a Zintl phase with incomplete
charge transfer from nominally divalent Ba.
However, the ionic character of Ba 
is less clear than that of the
alkali metals, as there is evidence for 
mixing of the Ba 5$d$ orbitals into the conduction
band.\cite{moriguchi00,saito95,blake01}
On the other hand, for the Cu-Ge clathrate studied here,
the valence counting argument works very well.
If we assume Cu to have valence 1
and Ba valence 2, valence counting predicts the stable
phase to be Ba$_{8}$Cu$_{(16/3)}$Ge$_{(46-16/3)}$ = 
Ba$_{8}$Cu$_{5.3}$Ge$_{40.7}$. Or, if framework
vacancies are allowed one obtains compositions
such as Ba$_{8}$Cu$_{5}$Ge$_{40.75}$.  These values are 
quite close to those observed in our 
samples (Table~\ref{tab:table1}).  

The formation of vacancies is not necessary by this argument.
However, the presence of Cu causes the lattice parameter to
contract relative to the Ba-Ge clathrate.  
Since Cu cannot occupy all of the 6$c$ sites by valence counting, 
some Ge atoms must occupy this
site, having a neighbor distance smaller 
than that of
elemental Ge (Fig.~\ref{fig:fig2}). Framework vacancies may
allow this strain to be accommodated more easily.
With an increased Cu concentration, the valence count
can also be maintained by inducing Ba vacancies, as observed
in Cu6.  This attests to the strong stabilizing
influence of the valence count, 
in contrast to the situation in the analogous Ag and Au
clathrates.

There are other intermetallics in which the valence 
count influences the stability, although the Zintl 
concept would not normally apply.  These would include
a large class of semi-Heusler alloys
such as TiNiSn, an intermetallic semiconductor.\cite{tobola98} 
However, in that
case the adherence to valence count is not nearly so 
strong as in BaCuGe clathrate. What distinguishes the
present case from other intermetallics is the fourfold-
coordinated framework which enhances the stabilization
of a local bonding configuration. This implies a rather
ionic configuration for Cu, which would be more difficult
to maintain for Au, and which may explain the distinction
between the Cu and Au clathrate.

The Zintl mechanism implies a deep minimum or gap
in the electron density of states at the Fermi level, which
is consistent with the large diamagnetic susceptibility
that we have measured. This susceptibility is four times 
larger than observed for the Au clathrate,\cite{herrmann99}
and larger than of other group-IV clathrates of which we are aware.
Per framework site, the susceptibility of the Cu6 sample is 
$-95 \times 10^{-6}$ emu/mol, comparable to that of the largest 
elemental diamagnets,\cite{gray72} and considerably larger 
than the value ($-13 \times 10^{-6}$ emu/mol or
$-1.4 \times 10^{-7}$ emu/g) we obtain for the core
susceptibility using standard parameters.\cite{mulay76}

In tetrahedrally-bonded elemental semiconductors, 
there is
a near-cancellation of the paramagnetic and diamagnetic
susceptibilities.  A model for this behavior\cite{hudgens74} 
indicates that the valence contribution to the diamagnetic 
susceptibility scales with
$\langle r^{2}\rangle$, as for molecular systems.
However, the expansion of the clathrate framework relative
to the zinc-blende lattice does not appear sufficient to
explain the large magnitude of the diamagnetic susceptibility 
observed here.  Furthermore, though there are structural
similarities to graphite, we do not expect the clathrates 
to have the conjugated orbitals which give ring currents
in aromatic molecules.\cite{mulay76}
On the other hand,
recent measurements\cite{mandrus98} have shown a large diamagnetic 
response for the semimetal RuAl$_{2}$, as well as other 
intermetallic semiconductor systems such as the skutterudite
CoSb$_{3}$. The mechanism for this is not known, however it
seems likely that the Cu clathrate is on the same footing with
these systems, and exhibits a narrow
gap or pseudogap at the Fermi surface.

\section{Conclusions}

We have prepared type-I clathrates from
different starting Ba-Cu-Ge compositions, and find 
a preferred composition with the Cu content
close to 5, rather than 6, per cell. Furthermore, the
composition adheres rather tightly to a valence-counting
scheme, which we attribute to the Zintl mechanism, in contrast 
to the Au and Ag analogs. This implies a semimetallic
or semiconducting characteristic, contrasting the 
reported metallic behavior of the Au clathrate.
A very large diamagnetic susceptibility was also observed,
which matches the behavior of other semiconducting and
semimetallic intermetallics.

\begin{acknowledgments}
This work was supported by the Robert A. Welch Foundation, 
Grant No. A-1526, and by Texas A\&M University through the
Telecommunications and Informatics Task Force.
\end{acknowledgments}

\bibliography{clathrateCux}

\end{document}